\documentclass[12pt]{article}
\usepackage{graphicx}
\begin{document}
\begin{center}
{\Large A Model of Varying Fine Structure Constant and Varying Speed of
Light}
\vskip 0.3 true in {\large J. W. Moffat}
\date{}
\vskip 0.3 true in {\it Department of Physics, University of Toronto,
Toronto, Ontario M5S 1A7, Canada}

\end{center}

\begin{abstract}%
The recent evidence for a cosmological evolution of the fine structure
constant $\alpha=e^2/\hbar c$ found from an analysis of absorption systems
in the spectra of distant quasars, is modelled by a cosmological scenario
in which it is assumed that only the speed of light varies. The model fits
the spectral line data and can also lead to a solution of the initial value
problems in cosmology.

\end{abstract} \vskip
0.2 true in
e-mail: moffat@medb.physics.utoronto.ca

\section{\bf Introduction}

The possibility that a large increase in the speed of light in the early
universe can solve the initial value problems in
cosmology~\cite{Moffat,Moffat2}, and present an alternative to the standard
inflationary models~\cite{Guth}, has received mounting
attention~\cite{Albrecht,Barrow,Barrow2,Brandenberger,
Martins,Harko,Magueijo,Clayton,Drummond,Liberati}. Higher-dimensional
models lead to a varying speed of light (VSL) when the radion field in
e.g. five-dimensional models varies~\cite{Kiritsis}. The recent increasing
evidence for a cosmological evolution of the fine structure constant in a
red shift range $0.5 \leq z\leq 3.5$ is therefore of considerable
interest~\cite{Webb}. The idea that the fine structure constant is varying
over the history of the universe has a long
history~\cite{Dirac,Teller,Dicke,Gamow,Dyson,Bekenstein}, and if
this discovery is confirmed by further observations and analyses, then it
will have a profound impact on the future of physics. Models of varying
$\alpha$ have been proposed
recently~\cite{Barrow3,Sandvik,Dvali,Youm} and a study has been
made of the possibility that the cosmic microwave background (CMB) data
could be used to detect a variation in $\alpha$~\cite{Martins2}. It is an
old argument~\cite{Dicke,Bekenstein2} that observations cannot measure
directly fundamental dimensional constants. Only dimensionless constants
such as the fine structure constant can be measured, which involves the
dimensional constants $e$, $c$ and $\hbar$. However, we can form
theoretical prejudices about which dimensional constants are responsible
for a variation of dimensionless fundamental constants and this entails
different models that describe the variation of these constants. Moreover,
independent observational evidence can be obtained that can rule out one or
another of these models.

In the following, we shall study a simple model that incorporates a VSL
behaviour in a cosmological setting, assuming that the electric charge $e$
and Planck's constant $\hbar$ are truly constants of nature. A detailed
analysis of a possible variation in the electric charge $e$ was given by
Bekenstein~\cite{Bekenstein}, in which the local gauge invariance of the
electromagnetic field was preserved, although conservation of charge was
broken. More recently,
Bekenstein's model has been reanalyzed in a cosmological
setting by Sandvik, Barrow and Magueijo~\cite{Sandvik}. One of Bekenstein's
conclusions, based on reasonable physical assumptions, was that spatial
gradients of the electric charge would cause a large discrepancy with the
weak equivalence principle experiments~\cite{Will}. If such a violation
of the weak equivalence principle is extrapolated back to red shifts in
the early matter dominated era of the universe, when the density of matter
was greater, then such spatial gradient violations could be so
large as to imply a significant violation of the weak equivalence
principle in that era. Variations of Planck's constant $\hbar$
at red shifts of order $z\sim 2-3$ could significantly affect
atomic and molecular spectral line observations and other
quantum phenomena. In view of this, we find that the possibility
of a varying speed of light is more attractive, even though our
understanding of special relativity, general relativity and
spacetime will be significantly altered. We should also
emphasize that a VSL explanation of a varying fine structure
constant has the important theoretical consequence of being able
to resolve the initial value problems of cosmology, whereas at
this stage of our theoretical understanding, it is not clear
what advantages there could be for a varying $e$
or $\hbar$.

In Maxwell's theory, the speed of light is predicted from the equation
\begin{equation}
c=\frac{1}{\sqrt{\epsilon\mu}},
\end{equation}
where $\epsilon$ and $\mu$ denote the electric permittivity and the
magnetic permeability of the vacuum, respectively. If we have a varying
speed of light, then we can write
\begin{equation}
\label{vacuumchi}
c(x)=\frac{1}{\sqrt{\chi(x)}},
\end{equation}
where $\chi$ is a function of the spacetime coordinates. This implies that
we picture the vacuum as a variable medium and the velocity of
electromagnetic waves depends on the magnitude of $\chi$. In
particular, the increase in the value of $c$ in the
early universe would be traced to a phase transition in the function
$\chi$, associated with a spontaneous symmetry breaking of
Lorentz invariance of the vacuum~\cite{Moffat}.

Once $\chi$ is treated as a function of the spacetime coordinates,
then we can no longer simply change units such that $\chi=1$.

\section{\bf Varying Fine Structure Constant and Speed of Light Model}

We shall use a simple minimal scheme
to illustrate physical consequences of a VSL, and defer the investigation
of a more geometrically rigorous theory of VSL, such as the bimetric
theory~\cite{Clayton,Drummond} to a future publication. In a
minimally coupled VSL theory, one replaces $c$ by a field in a preferred
frame of reference, $c(t)=c_0\phi(t)$, where $c_0$ denotes the present
value of the speed of light. The dynamical variables in the
action are the metric $g_{\mu\nu}$, matter variables
contained in the matter action, and the scalar
field $\phi$ which is assumed not to couple to the metric
explicitly~\cite{Albrecht,Barrow}. In the preferred frame the
curvature tensor is to be calculated from $g_{\mu\nu}$ at
constant $\phi$ in the normal manner. Varying the action with
respect to the metric gives the field equations
\begin{equation}
G_{\mu\nu}-g_{\mu\nu}\Lambda=\frac{8\pi G}{c_0^4\phi^4}T_{\mu\nu},
\end{equation}
where $G_{\mu\nu}=R_{\mu\nu}-\frac{1}{2}g_{\mu\nu}R$, $\Lambda$ is the
cosmological constant and $T_{\mu\nu}$ denotes the energy-momentum
tensor. This theory is not locally Lorentz invariant. Choosing a specific
time to be the comoving proper time, and assuming that the universe is
spatially homogeneous and isotropic, so that $c$ only depends on time
$c=c(t)$, then the FRW metric can still be written as
\begin{equation}
ds^2=c^2dt^2-a^2\biggl(\frac{dr^2}{1-kr^2}+r^2d\Omega^2\biggr),
\end{equation}
where $k=0,+1,-1$ for spatially flat, closed and open
universes, respectively. The Einstein equations are still of the form
\begin{equation}
\biggl(\frac{{\dot a}}{a}\biggr)^2=\frac{8\pi
G}{3}\rho_m-\frac{kc^2}{a^2}+\frac{c^2\Lambda}{3},
\end{equation}
\begin{equation}
\label{accelerationeq} \frac{{\ddot a}}{a}=-\frac{4\pi
G}{3}\biggl(\rho_m+3\frac{p_m}{c^2}\biggr)+\frac{c^2\Lambda}{3}.
\end{equation}
The conservation equations are modified due to the time dependence of $c$
(we assume that $G$ is constant):
\begin{equation}
\label{conservationeq} {\dot\rho}+3\frac{{\dot
a}}{a}\biggl(\rho_m+\frac{p_m}{c^2}\biggr) =\frac{3kc^2}{4\pi Ga^2}
\frac{{\dot c}}{c},
\end{equation}
where $\rho=\rho_m+\rho_\Lambda$ and $\rho_\Lambda=c^2\Lambda/8\pi G$.
We can fit the present cosmological data by choosing $\Omega_{0m}=0.3$ and
$\Omega_{0\Lambda}=0.7$, where $\Omega_{0m}=8\pi G\rho_m/3H_0^2$ and
$\Omega_{0\Lambda}=8\pi G\rho_\Lambda/3H_0^2$~\cite{Perlmutter}.

As shown in
ref.~\cite{Moffat}, and subsequently in
refs.~\cite{Albrecht,Barrow,Clayton}, VSL theories can solve the horizon,
flatness and particle relic problems of early universe cosmology, when
$\phi(t)$ takes on large values in the very early universe. The basic
problem of the existence of cosmological horizons in the standard big bang
model, leads to the puzzle that regions that have not be in causal contact
have the same physical properties. This puzzle is solved in VSL theories
by considering the proper distance
\begin{equation}
d_H=a(t)\int_{t_0}^t\frac{dt'c(t')}{a(t')}.
\end{equation}
For a large increase in the value of $c(t)$ corresponding to light
travelling faster in the early universe,
it is possible for the horizon to be much larger, so that all regions
in our past have been in causal contact. The flatness problem is also
explained, for if the speed of light undergoes a sharp change in a phase
transition, then it can be shown that $\Omega-1\sim 0$ is an attractor
solution for $\vert \dot c/c\vert < 0$, i.e. the speed of light
decreases as the universe expands.

We model $\phi(t)$ by
\begin{equation}
\phi(t)=\frac{1}{1+A(t)\biggl[\biggl(\frac{t}{t_0}\biggr)^b-1\biggr]},
\end{equation}
where $t_0$ denotes the present age of the universe and $b$ is a
positive constant. The speed of light has the form
\begin{equation}
c(t)=\frac{c_0}{1+A(t)\biggl[\biggl(\frac{t}{t_0}\biggr)^b-1\biggr]},
\end{equation}
where $c(t_0)=c_0$ and for $t\rightarrow 0$, we have
\begin{equation}
c(t)\rightarrow \frac{c_0}{1-A(t)}.
\end{equation}
The change in the speed of light $c$ is
\begin{equation}
\frac{\Delta c}{c}\equiv
\frac{c-c_0}{c_0}
=\frac{A\biggl[1-\biggl(\frac{t}{t_0}\biggr)^b\biggr]}
{1+A\biggl[\biggl(\frac{t}{t_0}\biggr)^b-1\biggr]}.
\end{equation}

We shall assume that $A(t)$ is a slowly varying function of $t$ as
$t\rightarrow 0$ until some critical time $t=t_c$, when $A(t)$ undergoes a
sharp increase to $A(t_c)\sim 1$, resulting in
a sudden increase in $c(t)$. This sharp increase in $c(t)$
corresponds to a phase transition in the function $\chi(t)$, in
Eq.(\ref{vacuumchi}), such that $\chi(t_c)\sim 0$.

Let us write
\begin{equation}
\alpha(t)\equiv \frac{e^2}{\hbar c_0\phi(t)}=\frac{\alpha_0}{\phi(t)},
\end{equation}
where we have assumed that $e$ and $\hbar$ are strictly constants of
nature, $\alpha_0$ denotes the present value of
$\alpha$ and $\alpha(t_0)=\alpha_0$. This yields the fractional
varying value for the fine structure constant
\begin{equation}
\frac{\Delta\alpha}{\alpha}\equiv \frac{\alpha-\alpha_0}{\alpha_0}
=A(t)\biggl[\biggl(\frac{t}{t_0}\biggr)^b-1\biggr].
\end{equation}
In order not to spoil the agreement of
the standard cosmological model with the CMB
data at the red shift $z\sim 1000$ and big
bang nucleosynthesis results, we assume that $t_c\ll t_{NS}$
where $t_{NS}$ denotes the time of nucleosynthesis at the red
shift $z\sim 10^9$. Moreover, we assume that $A(t_{\rm CMB})
\sim A(t_{NS}) < 10^{-3}$ where $t_{\rm CMB}$
denotes the time of CMB.
 
Assuming that $A\sim {\rm const.}$, the time variation of
$\alpha$ is given by
\begin{equation}
\frac{\dot\alpha}{\alpha}=\frac{bA\biggl(\frac{t}{t_0}\biggr)^{b-1}}
{t_0\biggl\{1+A\biggl[\biggl(\frac{t}{t_0}\biggr)^{b}-1\biggr]\biggr\}}.
\end{equation}
For $b=1.5, A=10^{-5},\,t/t_0=0.125$
corresponding to $z\sim 3$ and $t_0=13.9$ Gyr we get
\begin{equation}
\frac{{\dot\alpha}}{\alpha}=3.8\times 10^{-16}\,{\rm yr}^{-1}.
\end{equation}
We observe that $\Delta\alpha/\alpha\sim -\Delta c/c$ for $A\ll
1$ and at the phase transition $t=t_c$ $\Delta\alpha/\alpha\sim
-1$ so that $\alpha(t_c)\sim 0$.

In Fig. 1, we display a fit to the quasar spectral line data
of Webb et al.~\cite{Webb} for $b=1.5$ and $A=10^{-5}$. \vskip
0.3 true in \begin{center}
\includegraphics[width=3.5in,height=3.5in]{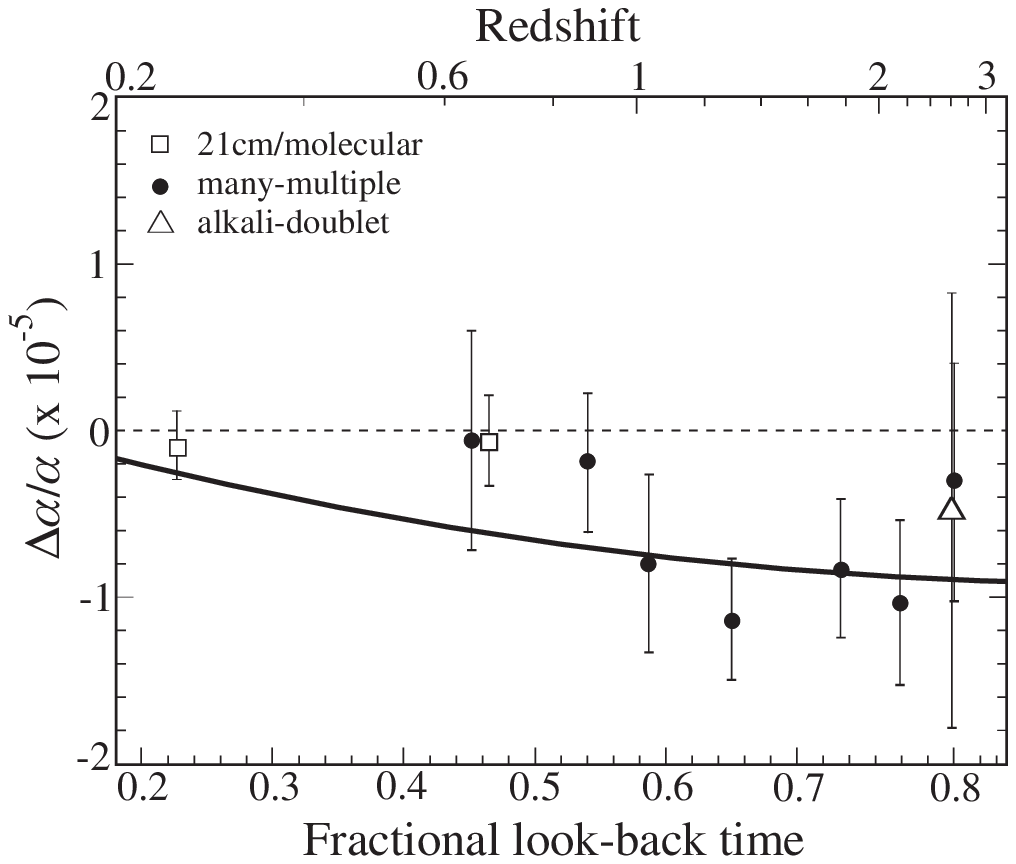}
\end{center} \vskip 0.1 true in \begin{center} Fig 1.
\end{center} \vskip 0.1 true in $\Delta\alpha/\alpha$ vs.
fractional look-back time to the big bang and red shift $z$. The
data points are from ref~\cite{Webb}. \vskip 0.1  true in

For $b=1.5$, $A=10^{-5}$ and $t/t_0=0.125$ ($z\sim 3$), we
get
\begin{equation}
\frac{\Delta c}{c}=0.9558\times 10^{-5},
\end{equation}
which is equivalent to a 1 part in $10^5$ increase in the presently
measured speed of light $c_0=299 792 458\,m\,s^{-1}$~\cite{Particle}.

\section{\bf Conclusions}
\vskip
0.3 true in

We have shown that a varying speed of light can explain the reported
variations in the fine structure constant, while satisfying all the
observational bounds. In particular, we can maintain the good agreement
with the CMB data and the big bang nucleosynthesis calculations.

Our simple model for a varying fine structure constant is phenomenological
in nature and serves the purpose of showing that a VSL model can be
consistent with the data, while resolving the initial value
problems in cosmology. A more fundamental model can be based on
a covariant bimetric formulation of speed of light and
gravitational wave speed, in which two light cones expand or
contract in the early universe~\cite{Clayton}. More research will
be needed to resolve the issue as to which fundamental constants
are varying and causing the reported variation in the fine
structure constant.

\vskip 0.3 true in
{\bf Acknowledgments}
\vskip 0.2 true in
I thank Pierre Savaria for a helpful discussion. This work was
supported by the Natural Sciences and Engineering Research
Council of Canada.
\vskip 0.5 true in

\end{document}